\documentclass[runningheads]{llncs}
\usepackage[T1]{fontenc}
\usepackage{amsmath}
\usepackage{graphicx}
\usepackage{subfig}
\usepackage{multirow,booktabs}
\begin{document}
%
\title{Social Isolation, Digital Connection:\\ COVID-19's Impact on Twitter Ego Networks}
\titlerunning{COVID-19's Impact on Twitter Ego Networks}
%
\author{Kamer Cekini\orcidID{0009-0004-7498-4986} \and
Elisabetta Biondi\orcidID{0000-0001-7464-1773} \and
Chiara Boldrini\orcidID{0000-0001-5080-8110}\and
Andrea Passarella\orcidID{0000-0002-1694-612X}\and
Marco Conti\orcidID{0000-0003-4097-4064}}
\authorrunning{K. Cekini et al.}
%
\institute{IIT-CNR, Pisa 56124, Italy \\
\email{k.cekini@studenti.unipi.it, \{elisabetta.biondi, chiara.boldrini, andrea.passarella, marco.conti\}@iit.cnr.it}
}
%
\maketitle              
%


\begin{abstract}
One of the most impactful measures to fight the COVID-19 pandemic in its early first years was the lockdown, implemented by governments to reduce physical contact among people and minimize opportunities for the virus to spread. As people were compelled to limit their physical interactions and stay at home, they turned to online social platforms to alleviate feelings of loneliness. Ego networks represent how people organize their relationships due to human cognitive constraints that impose limits on meaningful interactions among people. Physical contacts were disrupted during the lockdown, causing socialization to shift entirely online, leading to a shift in socialization into online platforms.  Our research aimed to investigate the impact of lockdown measures on online ego network structures potentially caused by the increase of cognitive expenses in online social networks. In particular, we examined a large Twitter dataset of users, covering 7 years of their activities. We found that during the lockdown, there was an increase in network sizes and a richer structure in social circles, with relationships becoming more intimate. Moreover, we observe that, after the lockdown measures were relaxed, these features returned to their pre-lockdown values. 

\keywords{ego networks  \and COVID-19 \and online social networks \and Twitter}
\end{abstract}
%
%
%

%
%
\section{Introduction}\label{sec:introduction}

The COVID-19 pandemic and subsequent lockdowns caused a profound change in societies around the world, with people being mandated to stay home for prolonged periods of time. This resulted in a seismic shift in which social interactions between people, almost overnight, moved from physical to digital spaces. All major social networking sites have experienced a never-seen-before boost of activities thanks to coronavirus lockdowns~\cite{ford_rojas_coronavirus_2020,schultz_keeping_2020}.
The activities of users on social networks, and specifically Twitter, during the COVID-19 pandemic have been extensively investigated~\cite{huang2020twitter,mattei2021italian,miyazaki2023fake}. However, there remains a notable gap in understanding how the unique context of lockdowns specifically altered the microcosms of social interactions on Twitter. Our work aims to address this gap by investigating the changes in online social interactions through the lens of \emph{ego networks}. 

\begin{figure}
    \centering
    \includegraphics[scale=0.15]{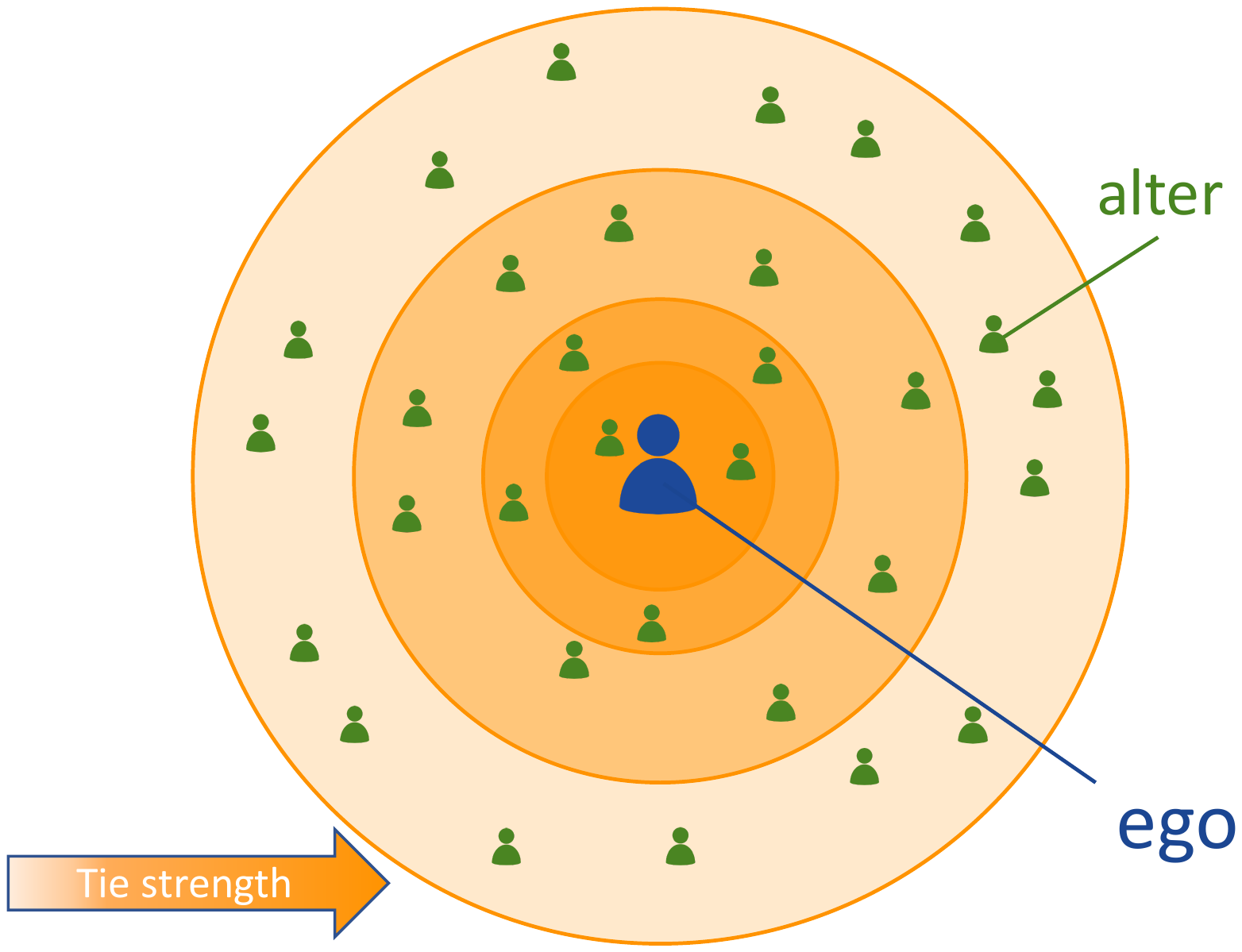}
    \caption{The ego network model.}
    \label{fig:egonet}
\end{figure}

The ego network model derived from evolutionary anthropology research on human social structures~\cite{Dunbar_1995}. It centers on a single individual, the \textit{ego}, and their immediate connections, the \textit{alters}. The model represents these relationships in concentric circles, with the closest and strongest ties to the ego in the smallest circle, and progressively weaker connections in larger circles (Figure~\ref{fig:egonet}). This structure reflects the varying degrees of intimacy in human relationships. Typically, ego networks consist of around 5, 15, 50, and 150 alters~\cite{Zhou_2005}, with a consistent size ratio of approximately 3 between each circle~\cite{Hill_2003}. Notably, ego networks only include relationships the ego actively maintains, representing those that are meaningful and nurtured over time. The relationships with alters not actively maintained are sometimes referred to as the inactive part of ego networks.

In this work, we aim to answer the following questions: (i) did lockdowns significantly alter the structure of user ego networks? (ii) did user activity and ego networks go back to their original level after lockdowns? To this end, we collected novel datasets comprising 1286 Twitter\footnote{Since our dataset was collected before Twitter changed its name to X, in this work we refer to the platform with its former name.} users, and we study how the ego networks of these users change over the years.
Our main findings are summarized below.
\begin{itemize}
    \item The active ego network size (i.e., the number of alters the ego spends cognitive effort to nurture) grows significantly after lockdown and returns to its normal size once lockdown restrictions are relaxed. This growth is statistically much higher than any growth seen in the five years prior to lockdown.
    \item As ego networks grow, they develop more circles, which again return to normal after lockdown. The circles that grow the most are the external ones, while the innermost circles remain stable (this is compatible with...).
    \item After lockdown, alters tend to move more towards inner circles, signifying a strengthening of relationships and increased intimacy during this period.
    \item After lockdown, egos gain several new alters and lose fewer than normal, but when lockdown restrictions are weakened, many alters are lost and fewer are gained. This signifies that egos have shifted their social cognitive capacity elsewhere (e.g., offline relationships).
\end{itemize}
In short, our results show that the ego networks of Twitter users grew significantly in the period after lockdown. Both the number of alters and the number of circles increased, indicating that users were able to allocate additional social cognitive capacity to online interactions during that period due to the lack of offline opportunities for socialization. However, this effect was temporary: as soon as restrictions were relaxed, the ego networks returned to their pre-pandemic status.

\section{Related work}
\label{sec:related_work}

Ego networks, as introduced in Section~\ref{sec:introduction}, model relationships between an individual (ego) and their peers (alters). This local subgraph, centered on the ego, uses tie strength (e.g., contact frequency~\cite{Hill_2003,roberts2011costs,arnaboldi2013ego}) to quantify relationship closeness.
Grouping ties by strength reveals \emph{intimacy layers} (Figure~\ref{fig:egonet}), typically sized 5, 15, 50, and 150, with decreasing closeness outwards. The outermost layer (150 alters) is known as \emph{Dunbar's number}, representing the maximum actively maintainable relationships~\cite{Hill_2003,Zhou_2005}.
This layered structure stems from limited cognitive capacity (the \emph{social brain hypothesis}~\cite{dunbar1998social}), leading to optimized resource allocation~\cite{sutcliffe2012relationships}. A key invariant is the \emph{scaling ratio} between layer sizes, often around three in offline and online networks~\cite{Dunbar2015,Zhou_2005}.
Dunbar's structure has been observed in various offline communication modes~\cite{roberts2009,Hill_2003,miritello2013} and online social networks (OSNs)~\cite{Dunbar2015}. This suggests OSNs are subject to the same cognitive constraints as traditional interactions.
Further research has explored tie strength and ego network formation~\cite{gonccalves2011modeling,quercia2012social}, their impact on information diffusion and diversity~\cite{aral2011diversity}, how ego networks and positive/negative relations are linked~\cite{tacchi2024keep}, and how ego networks can be applied effectively for link predition~\cite{toprak2022harnessing}.

%
%
\section{The dataset}
\label{sec:dataset}

Our research aims to identify how the COVID-19 pandemic and the lockdown measures implemented by governments have impacted the structure of online users' social networks. The pandemic originated in China in December 2019 and subsequently spread throughout the world at varying rates and times. Governments implemented diverse levels of lockdown measures, and the dates of implementation and relaxation varied depending on the evolving situation. For our analysis, we designated March 1, 2020, as the official start of the lockdowns (hereafter referred to as \emph{lockdown}), as this was when the Italian government initiated the first lockdown in the Western world. We then defined a time window of seven years, spanning from five years before the lockdown (March 1, 2016) to two years after (March 1, 2022). Within this time frame, we used a crawling method from the related literature~\cite{arnaboldi2013ego} to download the timelines of a large sample of over 10,500 user profiles. The crawling agent navigated the Twitter graph, with nodes representing users and edges indicating various forms of contact between users (such as followers, mentions, replies, or retweets). We initiated the crawling from Roberto Burioni's profile, an influential Italian virologist known for his campaign against anti-vaccination movements. When visiting a user, only a small and fixed number of neighbours are visited to maximise the distance from the seed and ensure a randomization of the sample.  For more information, please see~\cite{arnaboldi2013ego}. In the following, we will explain how we obtained the sample for our analysis from the provided dataset.


\subsection{Data cleaning and filtering}

We had to filter the data collected by the crawling agent for our specific needs. In the following paragraphs, we will explain how we eliminated non-human users and chose the ones suitable for the ego network analysis.

\subsubsection{Bot removal} 
Since our aim was to study human behaviour, we wanted to remove bots from our dataset. For this purpose, we use Botometer\footnote{\url{https://botometer.osome.iu.edu/}}, a web-based service that identifies bot accounts by looking at their features. Of the 53,837 users analyzed, we selected 10,547 non-bot users.

\subsubsection{Regular and active users}
In order to properly analyze ego networks on Twitter, users need to have consistent and active engagement on the platform, as mentioned in the relevant literature~\cite{arnaboldi2013ego,arnaboldi2017structure}. It is important for users to exhibit \emph{regular} and \emph{active} behaviour, with "regular" referring to how often they post and "active" reflecting their consistent use of the platform over the analyzed timeframe (please refer to the following for a detailed definition). The definition of regular and active users is time-sensitive, meaning that users who are consistent and active over a longer period may not be so when observed over shorter intervals. In the following, we will provide the definition of regular and active users $R_I$ over a generic period $I$.

\begin{description}
    \item[Active users in $I$] They are users who actively use Twitter during $I$. Following the definition given by~\cite{boldrini2018twitter}, for each user $i$ we compute the \emph{inactive life} $T_i^{inactive}$ as the length of the time interval between the last tweet and the end time of $I$. We compared this with the maximum time duration between two consecutive tweets $IIT_i$. If the inactive life is significantly longer than the intertweet time, specifically if $T_i^{inactive}>IIT_i + \textrm{ 6 months}$, then the user is considered to have ceased activity on the platform. 
    \item[Regular users in $I$] Following an approach used in~\cite{arnaboldi2017structure}, we also filtered sporadic users, whose activity is not robust enough to replicate realistic socialization interaction patterns. We identify \emph{regular} users based on how often they engaged in  \emph{social} interactions,  such as mentions, retweets and replies. We considered a user as regular if it participated in these interactions for at least 50\% of the months during the period $I$.  
\end{description}
Since our goal is to study the evolving ego networks year by year, we focus our research on users who are regular and active throughout each year. We define each time interval as $I_k=[01/03/2015+ k \textrm{years}, 01/03/2016+k \textrm{years}]$ for $k=0,\dots,6$, with the lockdown marking the boundary between $I_4$ and $I_5$. Thus, we keep only users that are regular and active in each interval, i.e.:
\begin{equation}
    R_0\cap R_1\cap R_2\cap R_3 \cap R_4 \cap R_5 \cap R_6.
\end{equation}

With this choice, we selected a very specific class of 1627 users, who are very social on Twitter. It is important to note that this group may not accurately represent all the typical regular and active Twitter users throughout the time window $I_0\cup\dots \cup I_6$. However, we state that this group constitutes a large and interesting sample of users whose characteristics may help identify some general trends in social habits. In the future, we plan to repeat our investigation for different classes of users.

\subsubsection{Outliers removal} 
\begin{figure}[t!]
     \centering
     \includegraphics[width=1.1\textwidth]{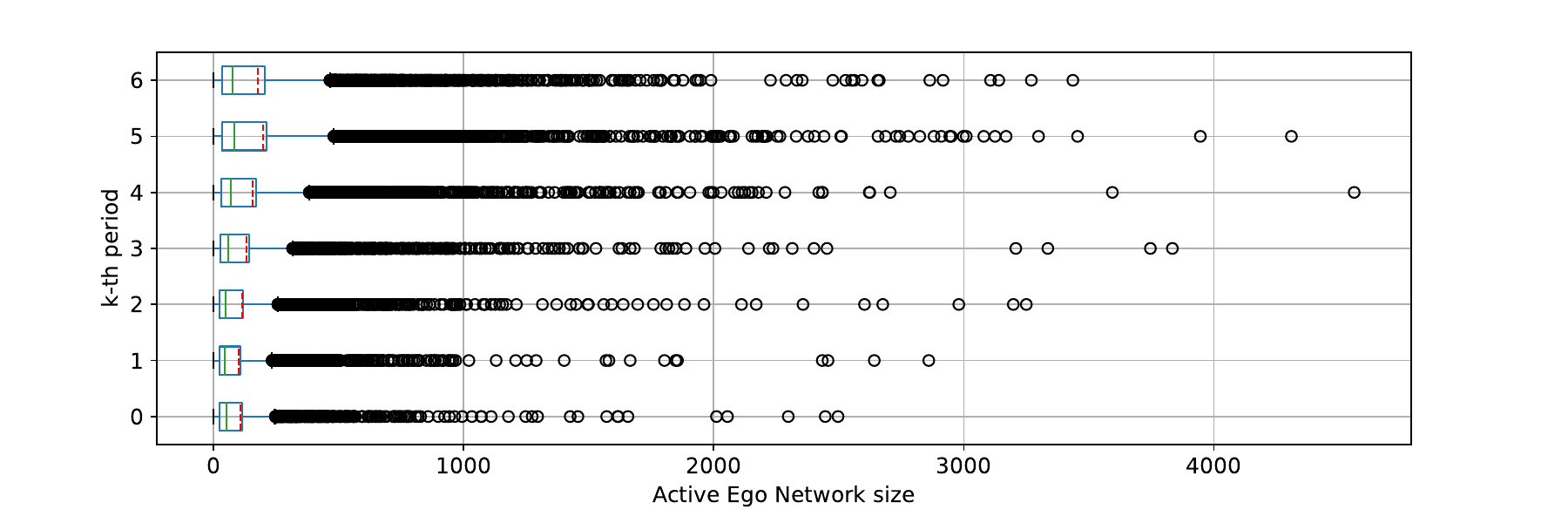}
    \caption{Distribution of active ego networks sizes}
    \label{fig:egonet_size_all}
\end{figure}

As can be seen in Figure~\ref{fig:egonet_size_all}, we observed great variability in active network sizes, with some users having active ego networks above 500, extremely large compared to the typical social features observed in the literature.   Because these users can significantly impact the overall statistics of the dataset, we have decided to treat them as outliers and remove them using the widely used interquartile range (IQR) method. As a result, we removed 341 outliers. 

\subsection{Dataset overview} 
\begin{figure}[t!]
     \centering
     \includegraphics[width=\textwidth]{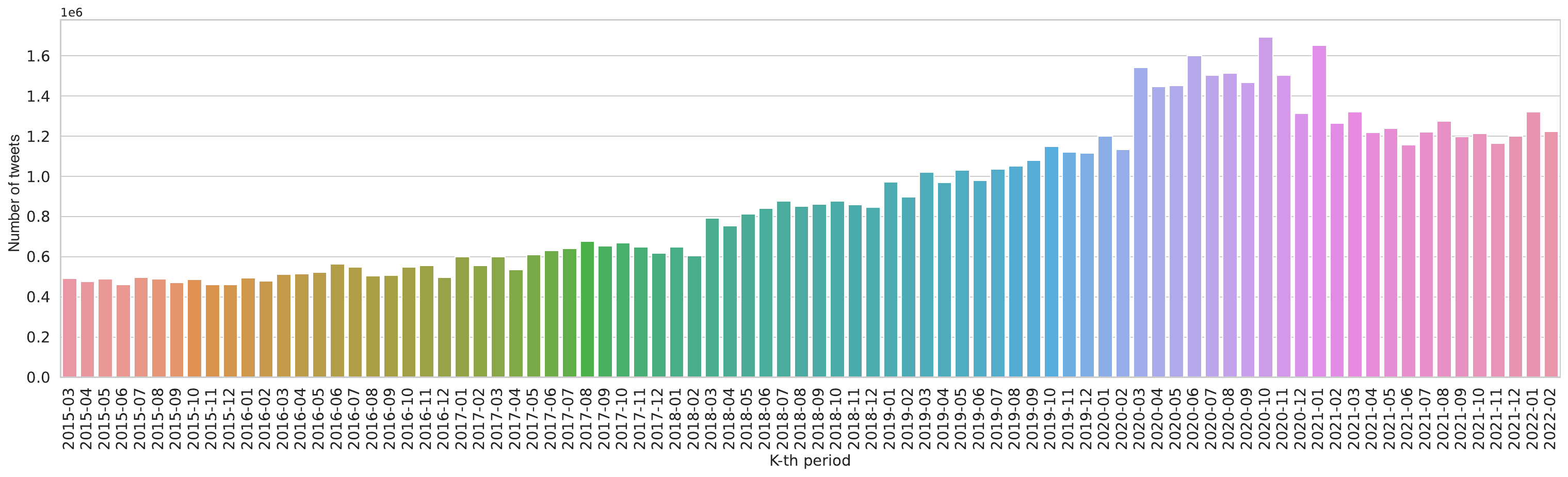}\vspace{-10pt}
    \caption{Total number of tweets considered}
    \label{fig:number_tweets}
\end{figure}

The dataset filtered with the approach described in the previous section is composed of 1,286 users. We ended up with a dataset consisting of over 67 million tweets. In Figure~\ref{fig:number_tweets}, we can see the distribution of the number of tweets over the time window and immediately observe a high peak in March 2020, corresponding to the lockdown.

\section{Methodology}
\label{sec:methods}

In this section, we briefly summarise how ego networks are constructed. 
The first step in computing ego networks is to calculate the ego-alter contact frequency. The contact frequency for online relationships is calculated as the number of direct interactions divided by the length of the relationships in years. We consider direct interactions replies, mentions, and retweets. Thus, the frequency of interactions (which is a proxy for social intimacy) between an ego $u$ and alter $j$ in period $I_k$ is given by the following formula:
\begin{equation}
w_{uj}^{(i)} = \frac{n_{reply}^{(u,j)} + n_{mention}^{(u,j)} + n_{retweet}^{(u,j)}}{I_i},
\end{equation}
where $n_*$ is the number of interactions from ego $u$ to alter $j$ and $I_i$ is the time length of the $i$-th period considered in our analysis. All the relationships with contact frequency $w_{uj}^{(i)} \geq 1$ are called \emph{active} and are part of the \emph{active} ego network of user $u$. 
After calculating the intimacy of the relationships as mentioned in the previous paragraph, we can group the active relationships into intimacy levels. To this aim, and similarly to \cite{boldrini2018twitter}, we use the Mean Shift algorithm. The advantage of Mean Shift, against, e.g., more traditional clustering methods like $k$-means, is that it automatically selects the optimal number of clusters. Each of the clusters found by Mean Shift corresponds to a \emph{ring} $\mathcal{R}$ in the ego network (Figure~\ref{fig:egonet}), with $\mathcal{R}_1$ being the one with the highest average contact frequency (i.e., intimacy). Then, circles $\mathcal{C}$ are obtained as the union set of concentric rings. Thus, it holds that $\mathcal{C}_{k} = \mathcal{C}_{k-1} \cup \mathcal{R}_k$, with initial condition $\mathcal{C}_1 = \mathcal{R}_1$. The active ego network size is thus the size of the largest circle. Note that we compute ego networks (hence their circles and rings) for each period $I_i$, so we will have circles $\mathcal{C}_{k}^{(i)}$ and rings $\mathcal{R}_k^{(i)}$.

In our analysis presented in Section~\ref{sec:results}, we focus on the following metrics: active ego network size (i.e., the number of alters contacted at least once a year), the optimal number of circles (as found by Mean Shift), and circle size (i.e., the number of alters in each circle of the ego network). These are the standard metrics for characterising ego networks. We compute one such value for each of the $i$ periods $I_i$ we consider. 
To provide statistical evidence for the observed trends, we will frequently use the concept of growth rate. Specifically, if $X_i^u$ indicates a quantity for a user $u$ during the year $I_i$, for example their active network size, we can look at its growth rate $G_{[i,i+1]}^u(X)$ between $I_i$ and $I_{i+1}$ defined as:
\begin{equation}
    G_{[i,i+1]}^u(X) = \frac{X^u_{i+1} - X^u_i}{X^u_i}.\label{eq:growth_rate}
\end{equation}
This rate is positive if there is an increase in the quantity or negative if there is a decrease.

\section{Ego network evolution during the COVID-19 Pandemic}
\label{sec:results}

In this section, we analyse the evolution of ego networks during the seven-year time window. Our analysis is twofold.  First, we investigate whether the decrease in in-person socializing due to lockdowns led to an increase in online activity within the network. We aim to uncover whether the cognitive effort previously directed towards offline interactions has shifted online.  Second, we examine the dynamic movement within the ego networks, delving deeper to understand the changes between different circles. 

\begin{figure}[t!]
     \centering
     \subfloat[Ego network sizes]{
         \includegraphics[width=0.48\textwidth]{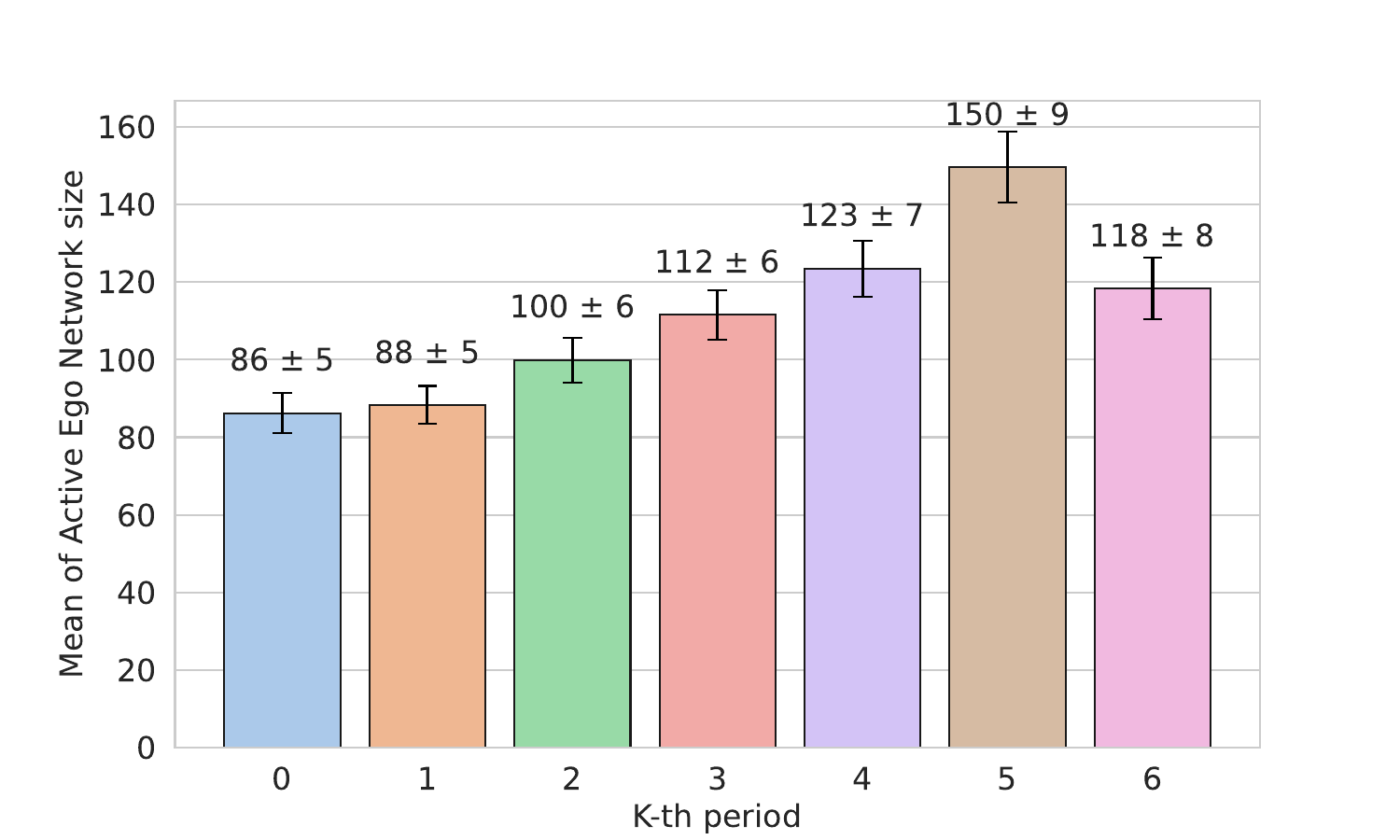}
     }\hfill
     \subfloat[Growth rates of difference of ego network sizes]{
         \includegraphics[width=0.48\textwidth]{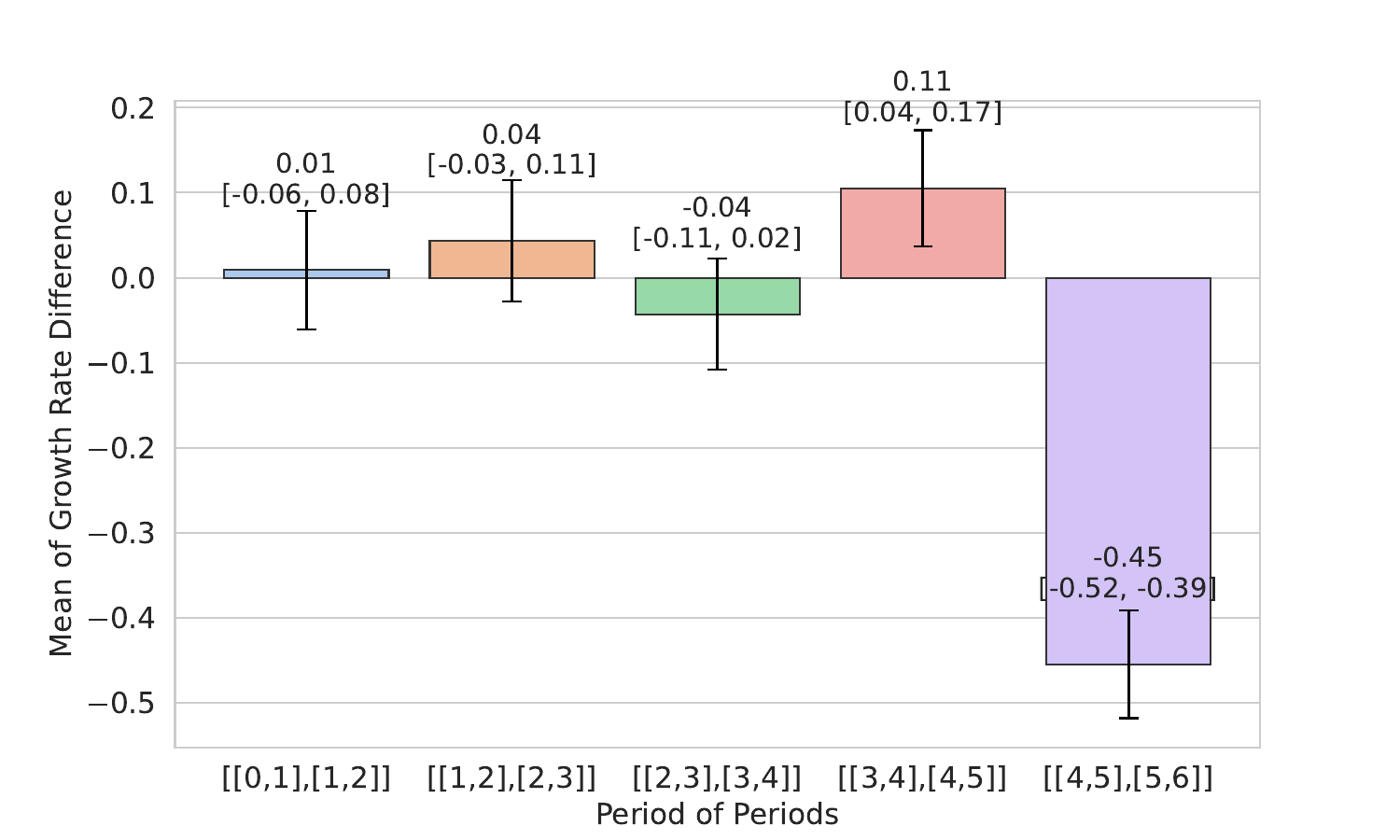}
     }\
        \caption{ Mean values and 99\% confidence interval of the ego network sizes in (a) and of the growth rate of their difference in (b).}
        \label{fig:egonet_size}
\end{figure}

We start our analysis by looking at the impact of reduced physical socialization on online ego networks. We will analyze the dimension of the active ego network size  $|A_i^u|$ for all users $u$ during the period $I_i$ and their growth rate\footnote{With $A_i^u$ we indicate the active network (i.e., the set of active alters), while with $|\cdot|$ indicate the cardinality of a set.}.
Figure~\ref{fig:egonet_size} shows the average values and the confidence interval of the active ego network sizes $|A_i^u|$ in (a) and their growth rates $G_{[i,i+1]}^u(|A|)$ in (b), for each period $I_i$. 
Two notable phenomena stand out. Firstly, there is a significant increase in ego network sizes in the year immediately following the lockdown, $I_5$, representing the highest increase in the entire time window (note also that the confidence interval in $I_5$ is not overlapping with those in the other years). Secondly, there is a decrease in size in the subsequent period, $I_6$, which is the only reduction in size observed in the dataset. These findings suggest that immediately after the lockdown, when many social interactions occurred exclusively online, users have more cognitive resources to invest in online socialization, leading to significantly larger ego networks. Conversely, the year after, when the more restrictive pandemic countermeasures were relaxed, online socialization returned to pre-lockdown levels. The data also revealed a growth in ego network size in the years leading up to the lockdown, possibly due to the rising popularity of the Twitter platform. However, the spike in size during the lockdown was notably higher than before. 

To support our claim, we examined the differences in active ego network size $(D_{|A|})_i=|A_{i}^u|-|A_{i-1}^u|$ between two consecutive time intervals and then looked at their growth rate, obtained as in Equation~\eqref{eq:growth_rate}. A positive growth rate $G_{[i,i+1]}^u(D_{|A|})$ expresses that $u$'s active ego network increases at a higher pace during $I_{i+1}$ than during $I_i$. We then conducted two t-tests on these growth rate distributions with the null hypothesis being ``the growth rate is non-positive'' and ``the growth rate is nonnegative'' respectively. Results can be seen in Table~\ref{table:ttest_diff_sizes}.  The first test rejected the non-positive hypothesis only for $R_4$ with a $p$-value equal to $0.0\mathrm{e}^{-4}$, proving that only in the period after lockdown we can statistically state that the size increased. The second test, instead, rejected the hypothesis on the last period with the same $p$-value, proving the evidence of a decrease in size in the year $I_6$. Instead, for the years $I_1,\dots, I_4$ we cannot state that there is statistical evidence of any increasing/decreasing trend.

\begin{table}[t!]
\caption{t-tests of the growth rate of the difference of active ego network sizes.}\label{table:ttest_diff_sizes}
\setlength{\tabcolsep}{10pt}
\centering
\begin{tabular}{c ll ll}
\toprule
\multirow{2}{*}{period pair} & \multicolumn{2}{c}{$H_0: \,G^u_{[i,i+1]}(D_{|A|})\leq 0$} & \multicolumn{2}{c}{$H_0: \,G^u_{[i,i+1]}(D_{|A|})\geq 0$}\\
\cmidrule(lr){2-3} \cmidrule(lr){4-5}
& results & $p$-value&results & $p$-value\\
\midrule
$(I_1,I_0)$ & ACCEPTED & $0.3696$ & ACCEPTED & $0.6304$ \\
$(I_2,I_1)$ & ACCEPTED & $0.0594$ & ACCEPTED & $0.9406$ \\
$(I_3,I_4)$ & ACCEPTED & $0.9538$ & ACCEPTED & $0.0462$ \\
$(I_4,I_5)$ & \textbf{REJECTED} & $0.0000$ & ACCEPTED & $1.0000$ \\
$(I_5,I_6)$ & ACCEPTED & $1.0000$ & \textbf{REJECTED} & $0.0000$ \\
\bottomrule
\end{tabular}
\end{table}

\begin{figure}[t!]
     \centering
     \includegraphics[width=\textwidth]{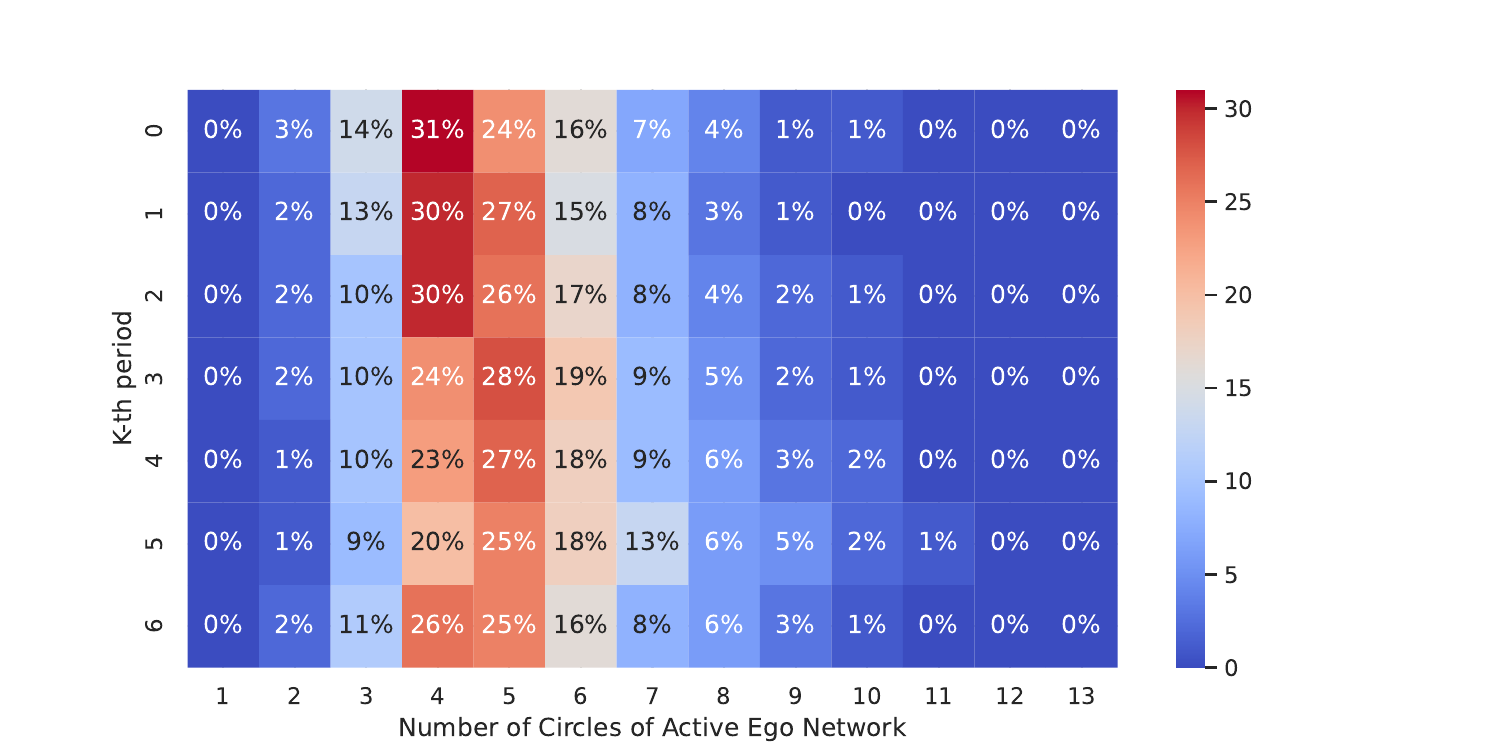}
    \caption{Distribution of ego networks' number of circles over the periods considered.}
    \label{fig:number_circles}
\end{figure}

We now examine the structure of ego networks over the years considered. Figure~\ref{fig:number_circles} shows the distribution of the number of circles in each time period. We can observe that the row corresponding to $I_5$, the year immediately after lockdown, is the one with the longest tail, with 13\% of users having a 7-circle ego network, and up to 1\% having a 11-circle ego network. The distributions corresponding to the rest of the periods are approximately similar, including that of $I_6$, suggesting that after the lockdown, the structure of ego networks returned to their previous patterns.

\begin{figure}[t!]
     \centering
     \includegraphics[width=\textwidth]{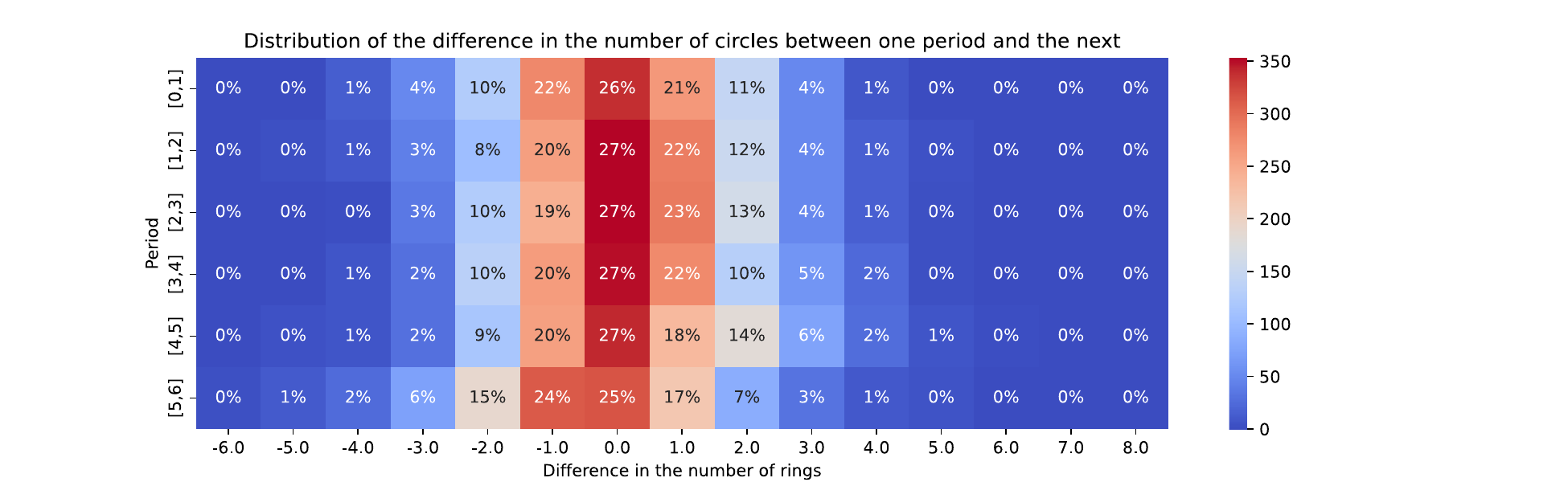}
     \caption{Distribution of the difference in the number of circles between two consecutive periods.}
     \label{fig:diff_numberCircles_heat}
\end{figure}

During the 7-year time window, the ego networks'  structures of circles vary over time, with many users changing the number of circles from one period to another. In Figure~\ref{fig:diff_numberCircles_heat} we plotted the distribution of the difference in the number of circles between two consecutive periods.
Figure~\ref{fig:diff_numberCircles_heat} shows that the most significant effects occur in the last two periods. In the second-last row, we observe, we can see that the weight of the right hand-
side of the distribution moves on the tail, indicating that users tend to expand their network structure with a higher number of circles compared to the previous periods.  Only $18\%$ users add a single circle, while 14\%, 6\%, 2\% and even a 1\% adding 2, 3, 4 and 5 circles respectively. In contrast, the left-hand side of the distribution remains similar to the previous periods.  In the last row of Figure~\ref{fig:diff_numberCircles_heat} the entire distribution shifts to the left, with the majority of nodes reducing the number of circles.  These results indicate that immediately after the lockdown, users' ego networks increased in size and became more intricate and complex. A year later, as the lockdown measures were relaxed, the opposite occurred, signifying a collapse in the structure of ego networks.

%

\begin{figure}[t!]
     \centering
    \subfloat[Egos with 5 circles]{
         \includegraphics[width=0.5\textwidth]{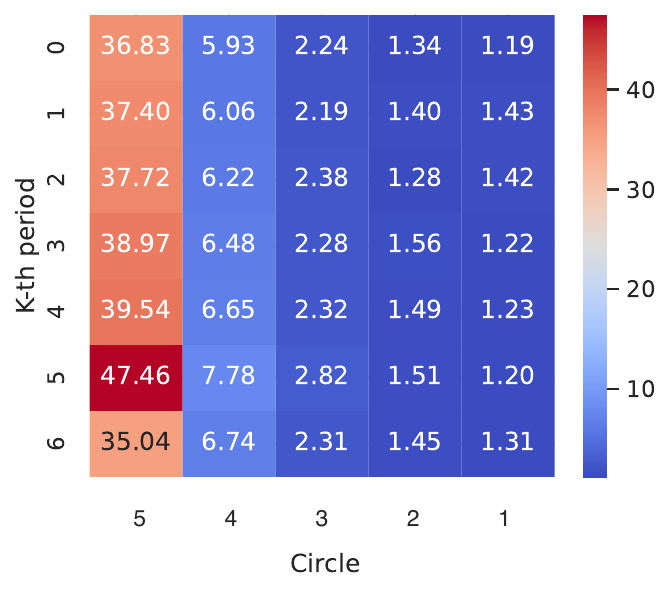}
     }\hfill
     \subfloat[Egos with 8 circles]{
         \includegraphics[width=0.7\textwidth]{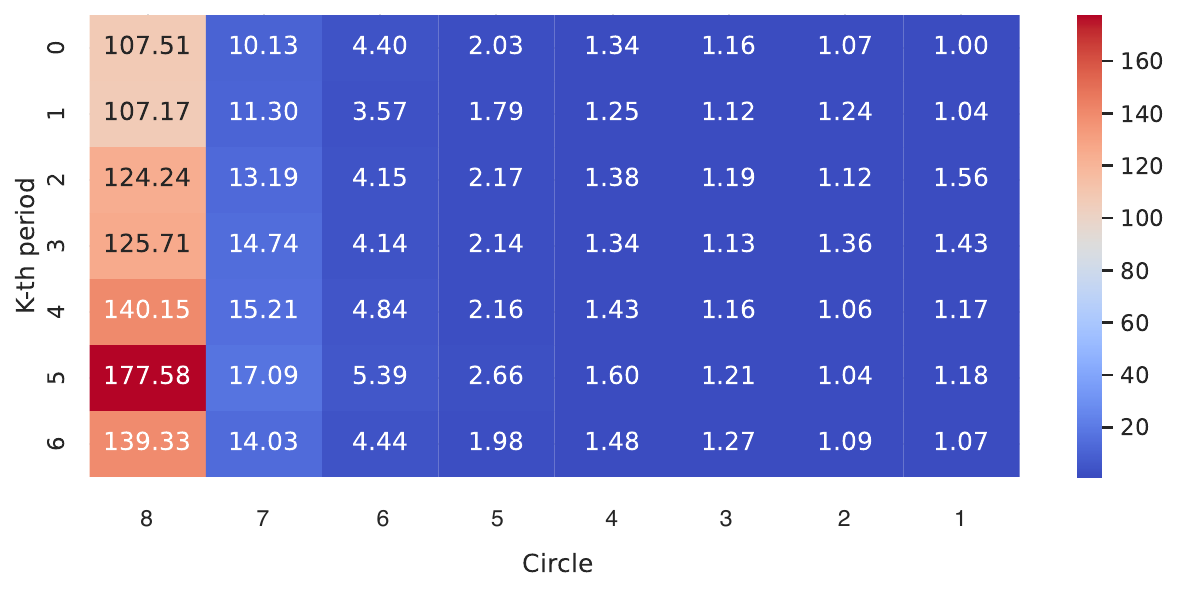}
     }\
        \caption{Dimensions of ego networks circles for egos that maintain 5 and 8 circles in their structure between two consecutive periods}
        \label{fig:dim_circles}
\end{figure}

We now focus on users whose ego networks exhibit the same number of circles between two consecutive periods. Figure~\ref{fig:dim_circles}.(a) shows the sizes of the circles or the users with respectively five circles in (a) and eight in (b),  within their ego networks. We can see that while inner circles maintain the same dimensions, the outer (circles 0 and 1) becomes larger in $I_5$, immediately after lockdown and then reduced again in $I_6$. We can also observe that the dimensions of the active ego networks (circle 0) are around 50/60 for egos with 5 circles and approximately 150/200 for egos with 8 circles, confirming that larger active circles reflect a more complex ego network structure.

%
%

\begin{figure}[t!]
     \centering
     \subfloat[]{
        \centering
         \includegraphics[width=0.4\textwidth]{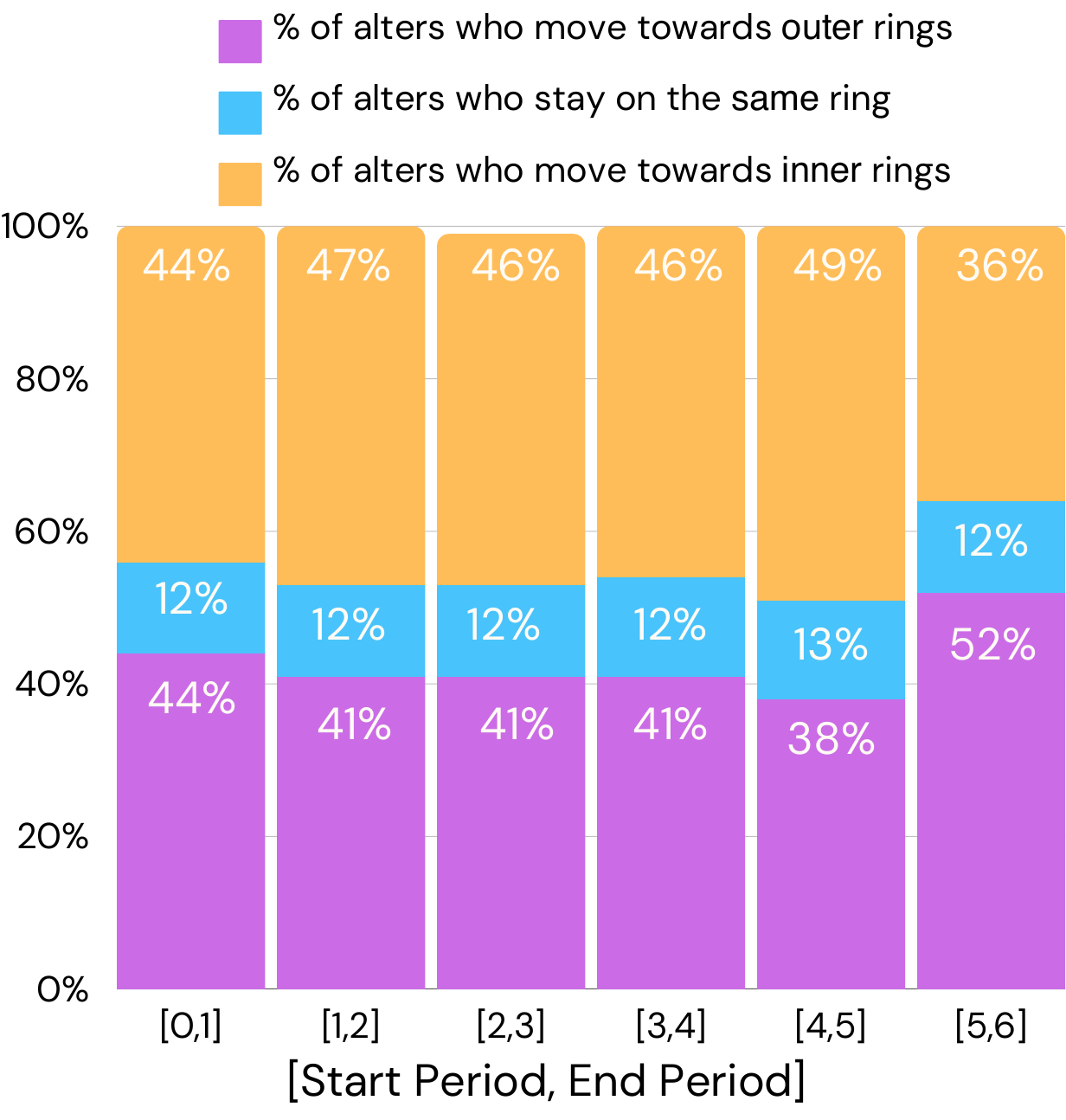}
     }\hfill
     \subfloat[]{
        \centering
         \includegraphics[width=0.4\textwidth]{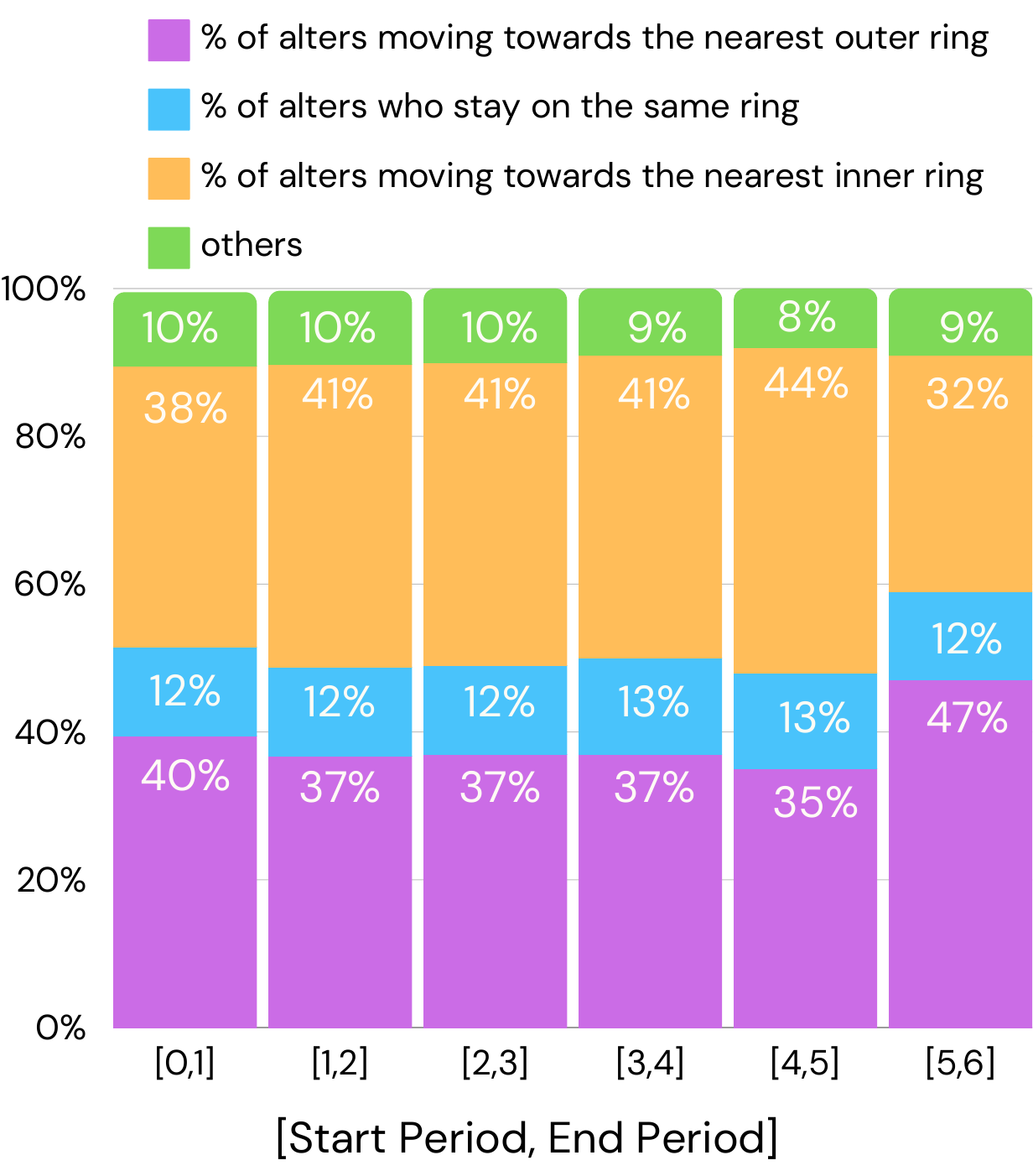}
     }\
        \caption{(a) Percentage of nodes moving toward inner, outer or remaining in the same ring. (b) Percentage of nodes moving toward the innermost, the outermost or remaining in the same ring.}
        \label{fig:movement_inside}
\end{figure}

The fluctuation in the number of circles suggests that individuals may be shifting between social circles over time. To investigate this further, we analyze the movement of users within these circles. Figure~\ref{fig:movement_inside} reveals that approximately half of the users (49\%) moved toward inner circles during period $I_5$, immediately following the lockdown. This signifies a strengthening of relationships and increased intimacy during this period.  Conversely, fewer users moved into outer circles during $I_5$ compared to previous periods. In contrast, during period $I_6$ (the second year after lockdown), the percentage of users moving into innermost circles reached its lowest point, while movement into outer circles peaked. This suggests a gradual return to pre-pandemic social patterns as relationships normalized.

\begin{figure}[t!]
     \centering
     \subfloat[Percentage of lost, stable and new alters]{
         \includegraphics[width=0.48\textwidth]{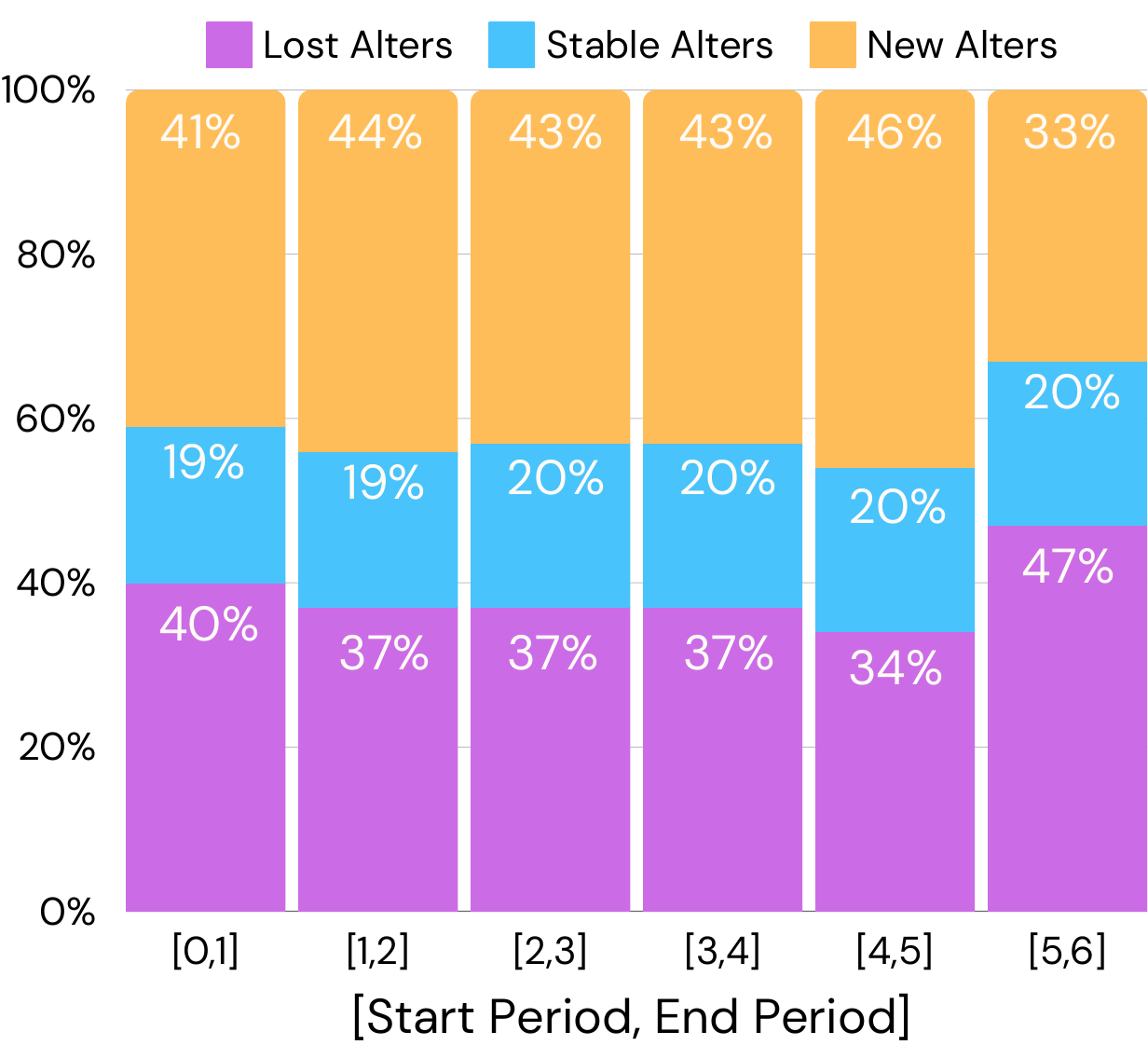}
     }\hfill
     \subfloat[Mean difference in the lost, stable and new alters's rates]{
         \includegraphics[width=0.48\textwidth]{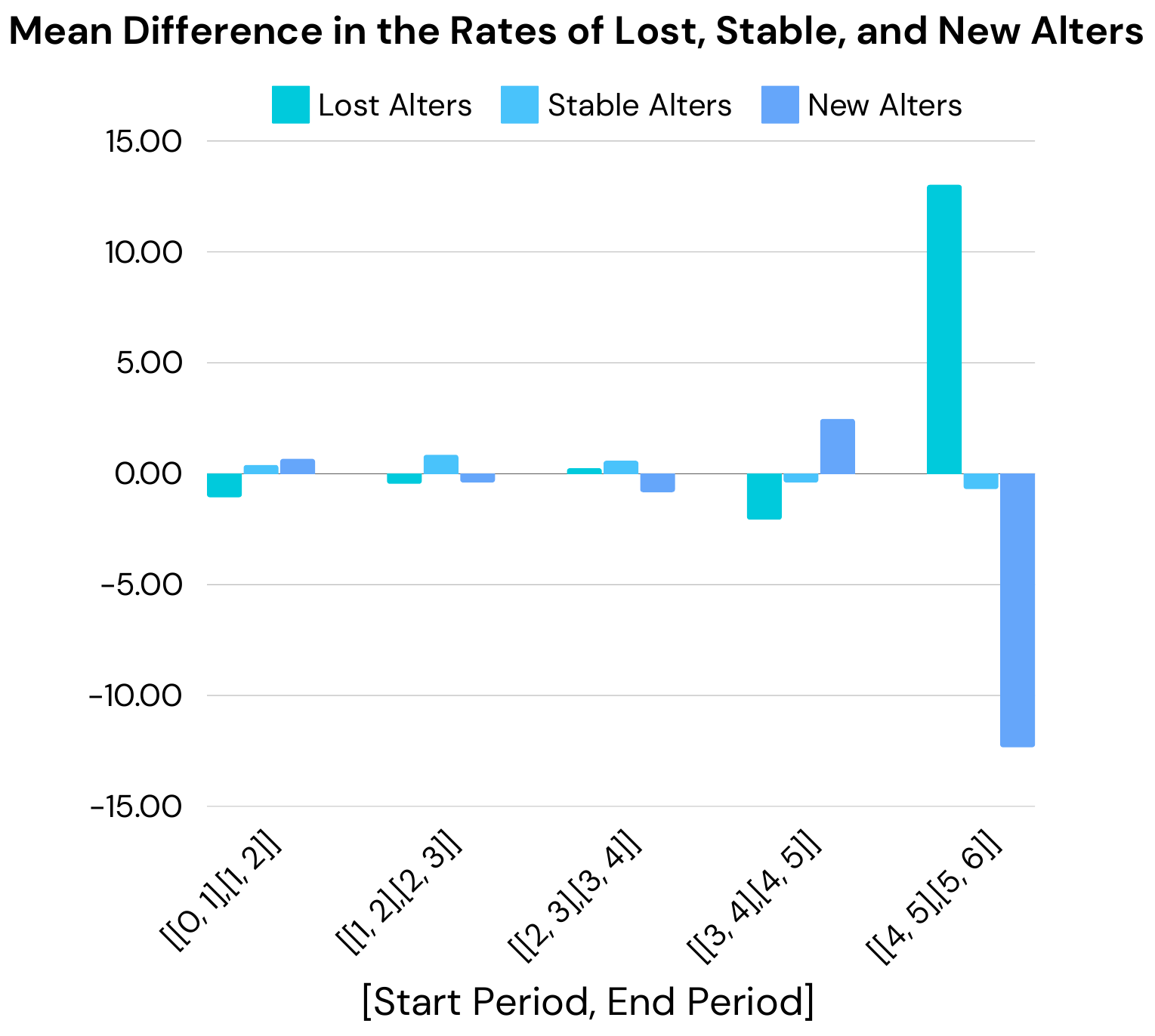}
     }\
        \caption{(a) Percentage of lost, stable and new alters between two consecutive periods (b) Mean difference in the rates of lost, stable and new alters}
        \label{fig:new_lost}
\end{figure}

We can now differentiate the nodes who enter, exit and remain within the ego network between two time intervals. 
For each pair of intervals $(I_i, I_{i+1})$, we will refer to the nodes that belong to active ego networks in both intervals as \emph{stable alters}. Nodes that did not belong to the ego network during $I_i$ but entered during $I_{i+1}$ are referred to as \emph{new alters}, while those who exited the ego networks (belonging to the ego network during $I_i$ but not during $I_{i+1}$) are called \emph{lost alters}. Lost, stable, and new alters between each interval pair $(I_i, I_{i+1})$ encompass the entire set of nodes in the active ego networks during $I_i$ and $I_{i+1}$, i.e., $A_i\cup A_{i+1}$. Figure~\ref{fig:new_lost}.(a) shows the percentage of them over the total number of alters in $I_i$ and $I_{i+1}$. We can observe that immediately after the lockdown there is an increase in the number of new alters entering the ego networks and a reduction in the number of those who leave them. In the following period, the opposite occurs. To validate these observations, we studied the distribution of growth rate in the difference in the fractions of lost, stable, and new alters for each ego between two consecutive periods. Specifically, after defining the quantities:
 \begin{align}
     \textbf{lost: }L_i^u &= \frac{|A_{i}^u\setminus A_{i+1}^u|}{|A_i^u\cup A_{i+1}^u|}\\
     \textbf{stable: }S_i^u &= \frac{|A_{i}^u\cap A_{i+1}^u|}{|A_i^u\cup A_{i+1}^u|}\\
     \textbf{new: }N_i^u &= \frac{|A_{i+1}^u\setminus A_{i}^u|}{|A_i^u\cup A_{i+1}^u|},
 \end{align}
we computed the differences $L_{i}^u-L_{i-1}^u$, $S_{i}^u-S_{i-1}^u$ and $N_{i}^u-N_{i-1}^u$ and then compute their growth rates as in Equation~\eqref{eq:growth_rate}. 
In Figure~\ref{fig:new_lost}.(b) are shown the mean values of these distributions 
and in Table~\ref{table:ttest_diff_perc} we show the outcomes of the t-tests.
The tests confirmed all the observations made above with a p-value smaller than $0.00$. Moreover, they also show an increase in stable alters in the two period triplets before lockdown and then a decrease. This is consistent with the fact that stable alters represent relationships that are supposed to be consolidated over time within the social network, which explains their initial growth. The lockdown, instead, represented a cutoff in the Twitter social network, which completely changed, with many people joining it for the first time or re-starting using an old account. 

\begin{table}
\caption{t-tests of the growth rate of the difference of the consecutive percentages of lost, stable and new alters.}\label{table:ttest_diff_perc}
\setlength{\tabcolsep}{10pt}
\centering
\begin{tabular}{c ll ll}
\multicolumn{5}{c}{Lost alters} \\ 
\toprule
\multirow{2}{*}{periods} & \multicolumn{2}{c}{$H_0: \,G^u_{[i,i+1]}(D_{L})\leq 0$} & \multicolumn{2}{c}{$H_0: \,G^u_{[i,i+1]}(D_{L})\geq 0$}\\
\cmidrule(lr){2-3} \cmidrule(lr){4-5}
& results & $p$-value&results & $p$-value\\
\midrule
$(I_2,I_1,I_0)$ & ACCEPTED & $0.93$ & ACCEPTED & $0.07$ \\
$(I_3,I_2,I_1)$ & ACCEPTED & $0.75$ & ACCEPTED & $0.25$ \\
$(I_4,I_3,I_2)$ & ACCEPTED & $0.35$ & ACCEPTED & $0.65$ \\
$(I_5,I_4,I_3)$ & ACCEPTED & $1.00$ & \textbf{REJECTED} & $0.00$ \\
$(I_6,I_5,I_4)$ & \textbf{REJECTED} & $0.00$ & ACCEPTED & $1.00$ \\
\bottomrule
\multicolumn{5}{c}{Stable alters} \\ 
\toprule
\multirow{2}{*}{periods} & \multicolumn{2}{c}{$H_0: \,G^u_{[i,i+1]}(D_{S})\leq 0$} & \multicolumn{2}{c}{$H_0: \,G^u_{[i,i+1]}(D_{S})\geq 0$}\\
\cmidrule(lr){2-3} \cmidrule(lr){4-5}
& results & $p$-value&results & $p$-value\\
\midrule
$(I_2,I_1,I_0)$ & ACCEPTED & $0.02$ & ACCEPTED & $0.98$ \\
$(I_3,I_2,I_1)$ & \textbf{REJECTED} & $0.00$ & ACCEPTED & $1.00$ \\
$(I_4,I_3,I_2)$ & \textbf{REJECTED} & $0.00$ & ACCEPTED & $1.00$ \\
$(I_5,I_4,I_3)$ & ACCEPTED & $0.99$ & \textbf{REJECTED} & $0.01$ \\
$(I_6,I_5,I_4) $ & ACCEPTED & $1.00$ & \textbf{REJECTED} & $0.00$ \\
\bottomrule
\multicolumn{5}{c}{New alters} \\ 
\toprule
\multirow{2}{*}{periods} & \multicolumn{2}{c}{$H_0: \,G^u_{[i,i+1]}(D_{N})\leq 0$} & \multicolumn{2}{c}{$H_0: \,G^u_{[i,i+1]}(D_{N})\geq 0$}\\
\cmidrule(lr){2-3} \cmidrule(lr){4-5}
& results & $p$-value&results & $p$-value\\
\midrule
$(I_2,I_1,I_0)$ & ACCEPTED & $0.19$ & ACCEPTED & $0.81$ \\
$(I_3,I_2,I_1)$ & ACCEPTED & $0.71$ & ACCEPTED & $0.29$ \\
$(I_4,I_3,I_2)$ & ACCEPTED & $0.89$ & ACCEPTED & $0.11$ \\
$(I_5,I_4,I_3)$ & \textbf{REJECTED} & $0.00$ & ACCEPTED & $1.00$ \\
$(I_6,I_5,I_4)$ & ACCEPTED & $1.00$ & \textbf{REJECTED} & $0.00$ \\
\bottomrule
\end{tabular}
\end{table}

\section{Conclusions}

In this work, we set out to investigate the effect of COVID-19 restrictions on the online socialization capacity of users on social networks. Specifically, we focused on the ego networks of these users, as ego networks are known to capture the structure and limits of human social cognitive efforts. We collected a novel dataset comprising the timelines of more than 1,000 Twitter users and extracted their ego networks over a seven-year period, including five years before the COVID-19 lockdown and two years after. Our main findings indicate that the active ego network size of Twitter users grew significantly after the lockdown, with a notable increase in the number of alters and circles, particularly in the external ones, while the innermost circles remained stable. Alters tended to move towards inner circles, signifying strengthened relationships and increased intimacy. Additionally, egos gained several new alters and lost fewer than normal during the lockdown, but many alters were lost and fewer were gained once restrictions were relaxed, suggesting a shift in social cognitive capacity towards offline relationships. In summary, our results show that ego networks expanded significantly during the lockdown due to increased online interactions, but this effect was temporary, with networks returning to their pre-pandemic status once restrictions were lifted.


\begin{credits}
\subsubsection{\ackname} This work was partially supported by SoBigData.it. SoBigData.it receives funding from European Union – NextGenerationEU – National Recovery and Resilience Plan (Piano Nazionale di Ripresa e Resilienza, PNRR) – Project: “SoBigData.it – Strengthening the Italian RI for Social Mining and Big Data Analytics” – Prot. IR0000013 – Avviso n. 3264 del 28/12/2021. 
C. Boldrini was also supported by PNRR - M4C2 - Investimento 1.4, Centro Nazionale CN00000013 - "ICSC - National Centre for HPC, Big Data and Quantum Computing" - Spoke 6, funded by the European Commission under the NextGeneration EU programme.
A. Passarella and M. Conti were also supported by the PNRR - M4C2 - Investimento 1.3, Partenariato Esteso PE00000013 - "FAIR", funded by the European Commission under the NextGeneration EU programme.

\subsubsection{\discintname}
The authors have no competing interests to declare that are relevant to the content of this article.

\end{credits}

%
%
\bibliographystyle{splncs04}
\bibliography{references}

\end{document}